\documentclass{aa}  
\usepackage{txfonts}
\usepackage{graphicx}
\begin{document}
   \title{A general method of estimating stellar astrophysical parameters from photometry}

   \author{
          A.N. Belikov\inst{1}$^{,}$\inst{2}
	  \and 
	  S. R\"oser\inst{2}
          }

   \offprints{A.N. Belikov,\\\email{A.N.Belikov@astro.rug.nl} }

   \institute{Kapteyn Astronomical Institute, P.O. Box 800, 9700 AV Groningen, the Netherlands \\
              \email{A.N.Belikov@astro.rug.nl}
             \and
             Astronomisches Rechen-Institut, M\"onchhofstrasse 12-14, D-69120, Heidelberg, Germany \\
             \email{roeser@ari.uni-heidelberg.de}
             }

   \date{Received 20 December 2007 / Accepted 21 July 2008 }

 
  \abstract
   {Applying photometric catalogs to the study of the population of the Galaxy is obscured by 
   the impossibility to map directly
   photometric colors into astrophysical parameters. Most of all-sky catalogs like ASCC 
   or 2MASS are based upon broad-band photometric systems, and the use of broad photometric bands 
   complicates the determination of 
   the astrophysical parameters for individual stars.}
   {This paper presents an algorithm for determining stellar astrophysical parameters (effective temperature, gravity and
   metallicity) from broad-band photometry even in the presence of interstellar reddening. This
   method suits the combination of narrow bands as well.}
   {We applied the method of interval-cluster analysis to finding stellar astrophysical parameters based on the newest Kurucz
   models calibrated with the use of a compiled catalog of stellar parameters.}
   {Our new method of determining astrophysical parameters allows all possible solutions to be located in the 
   effective temperature-gravity-metallicity space for the star and selection of  
   the most probable solution.}
   {}

   \authorrunning{Belikov \& R\"oser}\titlerunning{A General Method for APE from Photometry}
   
   \maketitle
%

\section{Introduction}

Estimating the astrophysical parameters of stars is one of the core problems in 
astrophysics 
and the key solving a number of astrophysical questions. Indeed, 
the knowledge of effective 
temperature, luminosity class (or gravity at the surface of the star), and 
metallicity of the star 
allows the distance to the star, the age of the star, and 
the extinction in the interstellar medium between the star and the observer
to be estimated. 

The number of stars with detailed spectral information is very limited
compared to the wealth of data in photometry. 
The problem of determining astrophysical parameters with photometry is usually 
solved with the help of 
the narrow-band photometric systems (for instance, the Str\"omgren photometric system) 
that provide quite good precision for a temperature, gravity, and
metallicity. For example, Nordstr\"om (\cite{Nordstroem}) estimates
the metallicity from 
Str\"omgren photometry   
with the final average 0.12 dex dispersion around the spectroscopic values. 
The problem becomes much more complicated in the case of broad-band
photometry where all fine features in the photometric bands (tracers of the 
differences in metallicity and especially in gravity)
are effectively smoothed. The estimation of $T_{eff}$, $\log g$, and [M/H] is crucial 
for the study of the Galaxy
on the basis of broad-band all-sky surveys like 2MASS or the broad-band deep surveys like
SDSS/SEGUE or UKIDSS.

The history of stellar parameter determination from broad-band photometry was started by
Johnson in the middle of the 50s of the last century (see, for example, 
Johnson~\cite{Johnson53} where the author 
defined extinction-free colors for 
the UBV photometric system).  Later, some broad-band
photometric systems showed the usefulness of extinction-free colors for determining 
of temperatures and
gravities with some assumptions about the metallicity (for example, the Vilnus photometric system).
The use of extinction-free colors is discussed in detail in Strai\v{z}ys (\cite{Straizys}). 

A major problem of using extinction-free colors is how to determine 
extinction-free colors themselves  from 
the observed ones. Indeed, as we show below, to calculate the extinction-free indices and to 
use them for the estimation of astrophysical parameters we have to accept some 
initial guesses about the estimated 
astrophysical parameters of the star {\it a priori}.

In this study we minimize  the amount of information to be assumed  {\it a priori} for 
the estimation of astrophysical parameters of the star. We will not make guesses about 
the astrophysical parameters of the star (temperature, gravity or metallicity) based on 
the position of the star, e.g. relative to the galactic plane. 

We do not discuss the dependence of the estimated parameter on the synthetic model grid adopted 
in this study and do not compare different synthetic model grids.
The estimated parameters will always depend on the consistency of the synthetic models used in the 
study. However, in the case of Johnson B,V,J,H, and $K_S$ photometry, we calibrated the parameters
obtained from the model fluxes by our method with published parameters from different authors.

\section{The synthetic grid of magnitudes and colors}

\begin{figure}
\centering
\includegraphics[width=9cm]{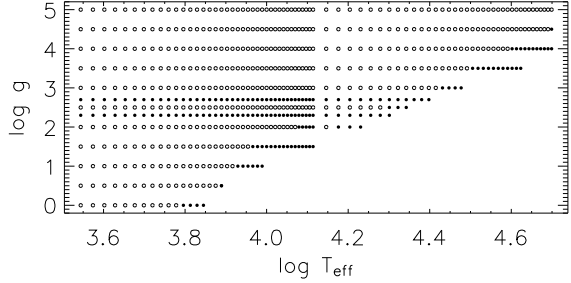}
   \caption{The synthetic grid of stellar atmospheres used in the study. Open circles are original models calculated by 
   Castelli (see Castelli \& Kurucz~\cite{Kurucz2006}) and the closed circles are models calculated by us with the use of ATLAS9 program 
   to fill the gap in the original grid. 
	   }
      \label{GridModels}
\end{figure}

To start with the determination of the astrophysical parameters from photometry we have
to provide a synthetic grid fine enough to give us synthetic magnitudes
and colors for each point of the grid. 

In this section we describe the definition of terms used in the paper, the computation 
of the synthetic grid of colors and magnitudes, 
the influence of extinction and the definition of the extinction-free colors. 
We create a grid of synthetic colors, use 
an assumed extinction law (Cardelli et al.~\cite{extinc}) to find reddened colors and to construct 
finally a grid of a synthetic extinction-free indices.  
\subsection{Definition of synthetic magnitudes}

The definition of synthetic magnitudes depends on the type of detector used in 
the observation that we are going to approximate. If the detected flux corrected for 
the atmospheric extinction is $f_{\lambda}$ and the stellar flux at the surface of the star 
is $F_{\lambda}$
(this is the flux delivered by synthetic atmosphere models), they are related to each other as 

\begin{equation}
f_{\lambda}=10^{-0.4\,A_{\lambda}}\,(R/d)^2\,F_{\lambda}
\end{equation}
where $A_{\lambda}$ is the interstellar extinction at the wavelength $\lambda$, 
$R$ the stellar radius, and $d$ the distance to the star, 
$R$ and $d$ are in the same units.
In the case of photon-counting devices the magnitude of the star $m_S$ in the filter $S_{\lambda}$ is
\begin{equation}
m_S =-2.5\,\log \frac{\int_{\lambda_1}^{\lambda_2} \lambda\,f_{\lambda}\,S_{\lambda} d\lambda}{\int_{\lambda_1}^{\lambda_2} \lambda\,S_{\lambda} d\lambda}
+m_S^0. 
\label{photon_count}
\end{equation}
We discuss the zero-points of
photometric systems $m_S^0$ in Sect.~\ref{calibration}.

\subsection{Theoretical models used in the study}

We used the latest Kurucz models of theoretical spectra (Castelli \& Kurucz~\cite{Kurucz2006}),
because these models can provide wider coverage in parameter space compared to other 
libraries. 
The grid used in the study consists of models with temperatures 
from 50,000 K to 3,500 K, $\log g$ from 0.0 to 5.0, and metallicities 
from [M/H]= --2.5 to 0.5. 
We extended this initial grid 
of models to lower gravity and added two points on the gravity axis 
($\log g=2.3$ and $\log g=2.7$)
using the program ATLAS9 by Castelli \& Kurucz (\cite{Kurucz2006}). The final 
grid of original models and newly calculated models is presented in Fig~\ref{GridModels}.  
These models do not have without overshooting, but do have microturbulent 
velocity $\xi = 2$ km/s.
The grid was interpolated with cubic splines to provide a step in $\Delta \log
T_{eff} = 0.005$ dex, a step $\Delta \log g =0.1$ dex, 
and a step $\Delta \left[\textrm{M/H}\right]=0.1$ dex. 

\begin{figure}
\centering
\includegraphics[width=9cm]{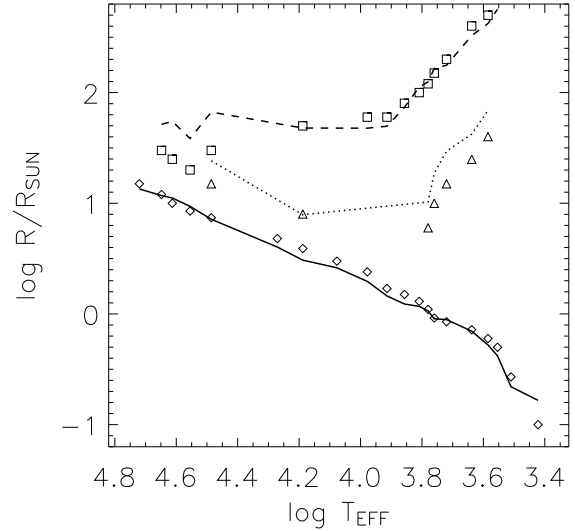}
   \caption{Comparison of radii used in this study with the radii for the main sequence (diamonds), 
   giants (triangles), and 
   supergiants (squares) from Schmidt-Kaler (\cite{LB}). The solid curve is the 
   theoretical radius-temperature dependence for the main sequence, 
   the dotted curve the theoretical radius-temperature dependence for giants, and 
   the dashed curve the theoretical radius-temperature 
   dependence for supergiants. 
	   }
      \label{FigRadii}
\end{figure}

To calculate absolute magnitudes from Kurucz models, we had to adopt stellar 
radii for each point in the $T_{eff} - \log g$ grid. We took radii 
from the theoretical track systems of Palla \& Stahler (\cite{Palla_Stahler}), 
D'Antona \& Mazzitelli (\cite{D'Antona}) for pre-main sequence stars, 
and  Geneva tracks (Schaller et al.~\cite{Maeder}) for post-main sequence stars. 
Figure~\ref{FigRadii} shows the agreement between the radii from the compiled track system (points) and the ``standard''
radii for main sequence stars, giants, and supergiants from Table~23 of Landolt-B\"ornstein 
(Schmidt-Kaler~\cite{LB}). The coincidence is generally quite satisfactory except in the case of
hot ($\log T_{eff} \ge 4.3$) supergiants.

We used the following sets of filters in the study: Johnson B and V filters from the 
UBVBUSER code (Buser~\cite{Buser}) and 2MASS J, H, and 
$K_S$ filters (Cutri et al.~\cite{2MASS}).

\subsection{Extinction in the case of a broad-band photometric system}

To compute a set of extinction-free indices, we have to compute first reddened magnitudes and 
colors for a wide range of 
interstellar extinction. This gives us synthetic color excesses for each model in the grid. 
We adopt the extinction law $A_{\lambda}$ from Cardelli et al.(\cite{extinc}), 
and take $A_V = 3.11\,E_{B-V}$, computing a grid of models with the step in extinction $\Delta A_V = 0.05\,\textrm{mag}$.

The extinction law used in the study (Cardelli et al.~\cite{extinc}) was constructed with the use of O- and early B-stars. This extinction law is 
a polynomial approximation of the extinction for the wavelength $\lambda$ $<A_{\lambda}/A_V> = a(1/\lambda)+b(1/\lambda)/R_V$.  
In fact, this value of 
$R_V$ is a parameter related to OB stars.   
This also means that extinction constructed with 
the use of  an extinction law $A_{\lambda}/A_V$ will give different values of the extinction 
for different spectral classes in the same 
photometric band and with the same distance and ISM between star and observer. 
Let us estimate this difference in extinction in the Johnson V band depending on the spectral class 
and luminosity class of the star. We calculate the extinction   
as the difference between theoretical magnitudes for the Johnson V filter and 
the stellar atmosphere
model with $\log T_{eff},\log g,[M/H]$ as  
$A_V^{model} = m_V(A_V^{input}) - m_V(A_V^{input}=0)$ with the use of $ A_{\lambda}(A_V) $
from Cardelli et al. (\cite{extinc}): 
\begin{eqnarray}
A_V^{model} = 
-2.5\,\log \frac{\int_{\lambda_1}^{\lambda_2}10^{-0.4\,A_{\lambda}(A_V^{input})}\lambda\,f_{\lambda}\,
S_{\lambda} d\lambda}{\int_{\lambda_1}^{\lambda_2} \lambda\,f_{\lambda}\,S_{\lambda} d\lambda}. 
\end{eqnarray}
The output $A_V^{model}$ differs from the input $A_V^{input}$ and this difference depends on 
the spectral class, gravity, and metallicity of 
the star. 
This difference is a physical phenomenon. Indeed, an O5 star and an M5 star being placed at 
the same distance 
from the observer and with the same optical depth of the ISM will be differently obscured by the ISM. 
This difference reaches up to 5\% of the nominal value of an extinction for the later spectral types. 

\begin{figure}
\centering
\includegraphics[width=9cm]{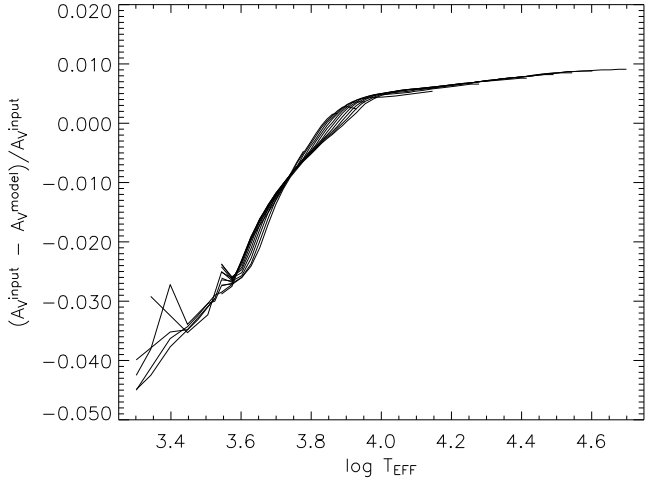}
   \caption{Differences in the extinction law. Solid lines are the differences in true and 
   nominal extinction in the Johnson V band for stars 
   of different temperatures for different gravities (from $\log g$ 0.0 to 5.0 with 0.5 dex step
   size). Solar metallicity is assumed.  
	   }
      \label{FigAvCorr}
\end{figure}

\subsection{Extinction-free color indices}

Extinction-free color indices (or Q-values) are the final target for our computations of 
synthetic colors. 
With the use of these indices, we
are able to reduce the number of models in our grid, because we 
can use a unique grid of synthetic 
models for any star independent of the individual extinction for this star. 

\begin{figure}
\centering
\includegraphics[width=9cm]{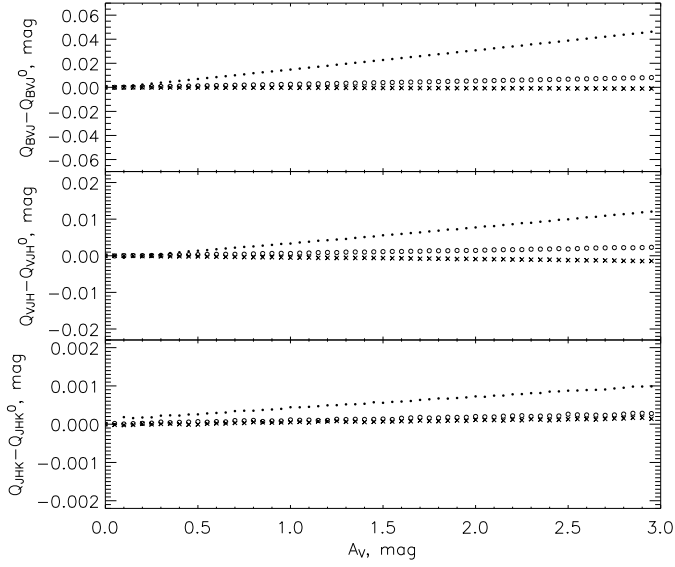}
   \caption{Dependence of the Q values on $A_V$ for stars with solar metallicities, 
   $\log g=4.5$, and $T_{eff}=3500$ K (filled circles), 
   $T_{eff}=7000$ K (open circles), and $T_{eff}=10000$ K (crosses).
	   }
      \label{FigQ_Av}
\end{figure}

\begin{figure}
\centering
\includegraphics[width=9cm]{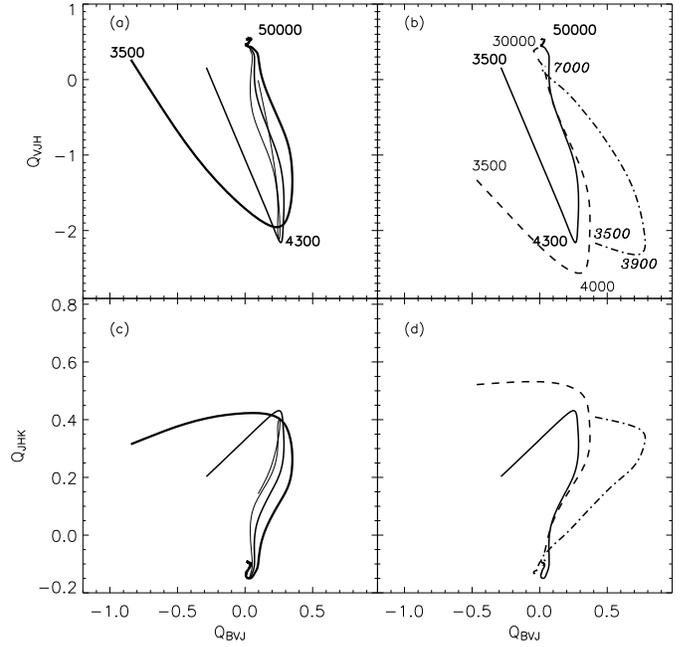}
   \caption{The Q diagrams. Panel (a) shows the $Q_{BVJ}$-$Q_{VJH}$ diagram for 3 different metallicities 
   (the thick line is for [M/H]=0.5, 
   the intermediate line for [M/H]=0.0, and the thin line for [M/H]=-0.5), and $\log g=4.5$. 
   Effective temperatures for [M/H]=0.5 are shown, effective temperatures for other metallicities 
   are, respectively, 50000 K for the point with maximum $Q_{VJH}$, 4300 K for the point with 
   minimum $Q_{VJH}$, and 3500 K for the end point of the sequence.  
   Panel (b) is the $Q_{BVJ}$-$Q_{VJH}$ diagram for solar metallicity and $\log g$ 0.0 
   (dotted-dashed line), 3.0 (dashed line), and 
   4.5 (solid line). Effective temperatures are shown for $\log g=0.0$ (in italics, 7000 K 
   for the point with maximum $Q_{VJH}$, 3900 K for the point with maximum $Q_{BVJ}$ and 3500 K 
   for the end point of the sequence), $\log g=3.0$ (30000 K for the point with maximum 
   $Q_{VJH}$, 4000 K for the point with minimum $Q_{VJH}$, and 3500 K for the end point of 
   the sequence),  $\log g=4.5$ (in bold, 50000 K for the point with maximum 
   $Q_{VJH}$, 4300 K for the point with minimum $Q_{VJH}$, and 3500 K for the end point of 
   the sequence). 
   Panels (c) and (d) are the same but for $Q_{BVJ}$-$Q_{JHK_S}$ diagram. 
	   }
      \label{FigQ_all}
\end{figure}

The construction of extinction-free color indices has some problems that we consider below. 
The Q-values are by definition 
\begin{eqnarray}
\label{Q_def}
Q_{m_1m_2m_3} & = (m_1-m_2)_0 - \frac{E_{m_1m_2}}{E_{m_2m_3}}\,(m_2-m_3)_0 \\
              & =(m_1-m_2) - \frac{E_{m_1m_2}}{E_{m_2m_3}}\,(m_2-m_3). \nonumber 
\end{eqnarray}
However, $E_{m_1m_2}/E_{m_2m_3}$ is not constant, but depends on the spectral class, gravity and metallicity of star
(see, for example, Grebel \& Roberts~\cite{Grebel95}). The excess ratio depends as well on the extinction itself. 
By definition, $E_{m_1m_2}= A_{m_1}-A_{m_2}$, so  
to remove extinction from the excess ratio, we have to replace $A_{m_1}$ with $A_{m_2}$ or 
vice versa. However, the 
dependence of the extinction on the optical depth is not linear 
(see, for instance, Chapter~6.5 from Strai\v{z}ys~\cite{Straizys}). 
As a result, we have a dependence $A_{m_1}=k_1\,A_{m_2}+k_2\,A_{m_2}^2+\dots$, and to construct 
an excess ratio, we have to neglect 
orders higher than the first in this dependence.  For the precise 
determination of astrophysical parameters, 
we have to check the possibility of neglecting these higher order coefficients. 
Figure~\ref{FigQ_Av} shows 
the remaining dependence of the Q-values on extinction, which can indeed be neglected compared to the Q-values 
themselves in Fig.~\ref{FigQ_all}.

\section{Calibration of theoretical models}
\label{calibration}

As we see from Eq.~(\ref{photon_count}), we have to bring our theoretical models into accordance 
with the zeropoint of the photometric system ($m_S^0$). 

To provide this calibration, we compiled a catalog of stars with temperature, gravity, and
metallicity known from spectroscopy. The catalog is a combination from the catalogs of Borkova \& Marsakov (\cite{borkova}) and 
Cayrel de Strobel et al. (\cite{cayrel}) with the addition of the 
catalogs of Soubiran el al. (\cite{soubiran}), Erspamer \& North (\cite{erspamer}) and Takeda et al. (\cite{takeda}). 
Our final list contains 3092 stars, 2096 of them were identified with the ASCC
catalog (Kharchenko~\cite{ASCC}). The latter contains 2,501,304 stars with Johnson B, V photometry recalculated from the Tycho $B_T$, $V_T$ and 
parallaxes and proper motions (Hipparcos and Tycho-1). We crossmatched the ASCC catalog with the 2MASS catalog, so that 
the stars in our compilation (2096 stars which we will call ``calibration catalog'' further on) have Johnson
$B$, $V$ and 2MASS $J$, $H$, $K_s$ magnitudes.  

A subsample of 1287 stars was selected with $\sigma_B < 0.1$ mag, $\sigma_V < 0.1$ mag, $\sigma_J < 0.3$ mag, $\sigma_H < 0.3$ mag, 
$\sigma_{K_s} < 0.3$ mag, parallaxes with $\sigma_{\pi}/\pi$ better than 10\% and distances within 100 pc. 
We assume that the stars from this subsample are extinction-free 
and use the subsample to calibrate our theoretical models.

To calibrate the theoretical sequences, 
we selected all stars from the subsample with $\log g = 4.0 \pm 0.1$, $[M/H] = 0.0 \pm 0.2$, 
and photometric errors in all bands better than 0.05 mag and absolute magnitudes in the B band $M_B < 6.0$ mag.    

On the basis of this calibration subsample of 58 selected stars, we calibrated the synthetic 
models 

\begin{eqnarray}
(B-V)^C = & (B-V)^T + (0.6001 \pm 0.0445) \nonumber \\
(V-J)^C = & (V-J)^T + (2.6764 \pm 0.0896) \\
(J-H)^C = & (J-H)^T + (1.0382 \pm 0.0361) \nonumber \\
(H-K_S)^C = & (H-K_S)^T + (1.1141 \pm 0.0290). \nonumber
\end{eqnarray} 
Although, the parameters from literature cover only a small fraction of the full parameter
space, we applied the corrections to the full sythetic grid.
Comparison of our calibrated colors with the colors for Vega shows good agreement. 
We take the atmospheric parameters for Vega from Smith \& Dworetsky (\cite{Vegaparam})  
($T_{eff}=9450 K$, $\log g=4.00$, [M/H]=-0.55), which 
from our calibrated model grid yields $(B-V)=0.005$ mag, $(V-J)=0.014$ mag, $(J-H)=-0.060$ mag, 
$(H-K_s)=0.052$ mag. The 2MASS $JHK_S$ photometric system differs slightly
from the Johnson $JHK$ (see Carpenter~\cite{carpenter} and Koen et al.~\cite{Koen}).

\section{The interval-cluster method of the astrophysical parameters estimation}

Parameters like effective temperature, gravity, metallicity, extinction, and distance 
on the basis of broad-band
photometry must be determined directly from photometry without any a priori assumption about physical
properties of stars that can restrict the
range of the search. 
Suppose that we have a parameter space 
\{$X_0$\}=\{$T_{eff}$,$\log g$, [M/H], $A_V$, r\}, where $T_{eff}$ is the effective temperature, 
$\log g$ the
logarithm of gravity, [M/H] the metallicity, $A_V$ the extinction in the Johnson V band 
($A_V$ is used as the parameter for extinction hereafter), and r the distance to the star. 
From observations (or from a synthetic grid in the case of simulations), 
we have a magnitude space \{$Y_0$\}=\{$m_1,\dots,m_n$\} 
that can be 
transformed to {\it the color space} \{ Y \}=\{ $m_i-m_j$, $i \ne j $\}. 
Via this transformation,  
we eliminate the distance r from the parameter space 
\{$X_0$\} and work in the parameter space \{$X$\}=\{$T_{eff}$,$\log g$, [M/H], $A_V$\}. 

Let us suppose that there are two different subsets in the parameter space 
\{ $X^1$ \} and \{ $X^2$\}. 
These two classes can be populated
by stars within any given temperature range, gravity range, and metallicity range. 
Let us as well have two subsets in the color space 
\{$Y^1$\} and \{$Y^2$\} containing the synthetic colors of the elements 
in  \{$X^1$\} and \{$X^2$\}. 
The task of classification is to attribute the star with some colors
to one of the classes in the parameter space \{X\}. This is possible if 
the two subsets \{$Y^1$\} and \{$Y^2$\} are separated in the
color space, but  
this requirement is not always guaranteed in the case of broad-band photometry. 
Stars with different temperatures, gravities, 
metallicities, and extinction can occupy the same area in the color space.

\begin{figure}
\centering
\includegraphics[width=9cm]{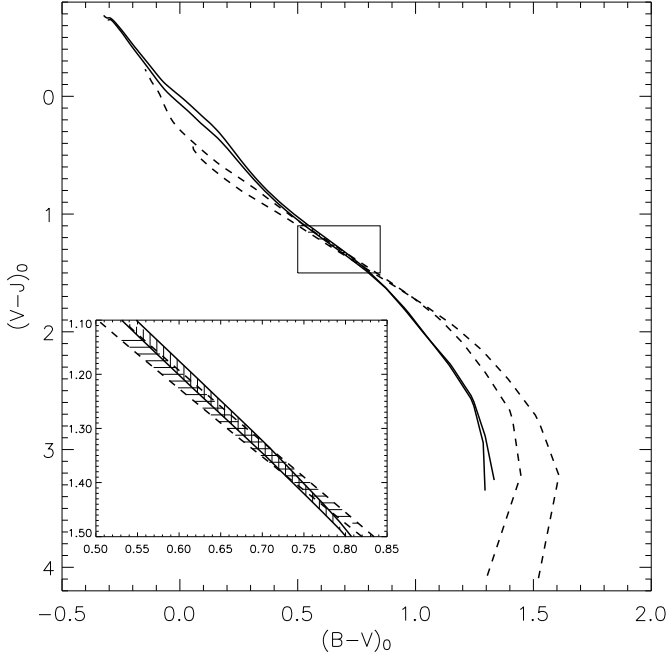}
   \caption{The color-color diagram for $(B-V)_0$ and $(V-J)_0$ colors. 
   Solid lines are for metallicity $[M/H]=-0.5$ and $\log g$ between 4.5 and 5.0, 
   dashed lines for solar metallicity and $\log g $ between 1.0 and 2.0.  
	   }
      \label{Fig_2c}
\end{figure}

Figure~\ref{Fig_2c} illustrates the problem: stars within two different 
metallicities and gravities ([M/H]=-0.5, $\log g \in [4.5,5.0]$ and 
[M/H]=0.0, $\log g \in [1.0,2.0]$) occupy the same space in the 2-color 
diagram even in the absence of extinction. 
It is obvious that for this range 
of colors there are at least two 
different solutions in parameter space for the star inside the filled area. 

In this case, there are at least two possible solutions for the parameters of a star: 
\{ $X^1 : [M/H]=-0.5, \log g \in [4.5,5.0]$ \} (dwarf) and \{$X^2:[M/H]=0.0, \log g \in [1.0,2.0] $\} (supergiant). 
The idea of the method we propose here is the location of all probable solutions  \{$X^i$\} and an 
estimation of the probability that the star can belong to the solution \{$X^i$\}.

To reduce the number of parameters in the parameter space, we transform the colors of 
the star into the extinction-free color 
indices described
above. The method is illustrated in the following 5 bands: Johnson B and V magnitudes (B,V filters are from UBVBUSER code, Buser~\cite{Buser}) 
and 2MASS J, H, and $K_S$ magnitudes (Cutri el al.~\cite{2MASS}). 
Four colors ($B-V$, $V-J$, $J-H$, $H-K_S$) and three extinction-free color indices 
($Q_{BVJ}$,$Q_{VJH}$, and $Q_{JHK_S}$) are used.

Our method consists of three steps: 
\begin{itemize}
\item[1.] interval analysis: select all theoretical models in our grid which satisfy the extinction-free color indices of the star;
\item[2.] cluster analysis: group all possible solutions in parameter space and find a set of solutions;
\item[3.] selection of the best solution: consider all possible solutions to select the only one as the most probable solution for 
the star in process.  
\end{itemize} 

The final solution for the star is a set of parameters (mean values and variances): 
\{$<T_{eff}>^i$, $<log g>^i$, $<[M/H]>^i$, $<A_V>^i$, 
$\sigma_{T_{eff}}^i$,$\sigma_{\log g}^i$,$\sigma_{[M/H]}^i$,$\sigma_{A_V}^i$\}, 
$i \in 1,n$, where n is the number of possible solutions for the star. 
One of these solutions will be selected as ``the most probable''
due to criteria we describe below. 

\subsection{The first part of the method: interval analysis}

\begin{table*}
\centering
\caption[]{The calibration subsample}
\label{table_sample}
\begin{tabular}{|c|c|c|c|c|}
\hline
Catalog & Number of &      \multicolumn{3}{|c|}{Interval in}          \\  
\cline{3-5}      
        &  stars   &  $ T_{eff}$, K & $\log g$, dex & [M/H],dex \\
\hline	
Borkova \& Marsakov (\cite{borkova})  & 876  & 4520 --- 6790  & 2.94 --- 5.00  & -3.85 --- 0.55 \\
Cayrel de Strobel et al. (\cite{cayrel})   & 3247 & 2291 --- 36000 & 0.09 --- 6.00  & -5.6  --- 2.9\\
Soubiran et al. (\cite{soubiran}) & 211  & 3922 --- 6276  & 0.58 --- 4.69  & -2.91 --- 0.34 \\
Erspamer \& North (\cite{erspamer})	& 140  & 5902 --- 10375 & 3.04 --- 4.36  & -1.08 --- 1.65  \\
Takeda et al. (\cite{takeda})   & 160  & 5009.3 --- 6967.7 & 3.185 ---  4.865  & -1.291 --- 0.457  \\
\hline
\end{tabular}
\end{table*}

In this step we locate possible solutions in parameter space ($\log T_{eff}$, $\log g$, [M/H]). 
The table of the theoretical models 
is constructed with steps $\Delta \log T_{eff} = 0.005$ dex, $\Delta \log g =0.1$ dex, 
and $\Delta \left[\textrm{M/H}\right]=0.1$ dex
and has 249922 entries in the parameter space. For each entry there is one 
combination of theoretical colors 
($(B-V)^T, (V-J)^T, (J-H)^T, (H-K_S)^T$) and 
extinction-free color indices ($(Q_{BVJ})_T$,$(Q_{VJH})_T$,$(Q_{JHK})_T$), as well as 
the set of supplementary 
parameters: ratios of extinction
($(A_B/A_V)^T$, $(A_J/A_V)^T$, $(A_H/A_V)^T$, $(A_{K_S}/A_V)^T$), 
ratios of reddening ($(E_{BV}/E_{VJ})^T$, $(E_{VJ}/E_{JH})^T$, 
$(E_{JH}/E_{HK_S})^T$). 
The ``observed'' values for the star are magnitudes ($B^S$,$V^S$,$J^S$,$H^S$,$K_S^S$) and colors
($(B-V)^S$, $(V-J)^S$, $(J-H)^S$, $(H-K)_S^S$).

\subsubsection{Selection of grid points}

The selection of grid points satisfying the set of ``observed'' colors is carried out in 
the space of extinction-free color indices. 
The value $(Q_{BVJ})^T$ does not depend on extinction. 
If we know stellar parameters ($\log T_{eff}, \log g, [M/H]$),
we can find a stellar excess ratio ${E_{BV}/E_{VJ}}^S$. Then we can construct 
an ``observed'' extinction-free color 
index $Q_{BVJ}^S$ and compare it with the theoretical ones $Q_{BVJ}^T$ 
to select grid points satisfying to the input index Q: 

\begin{equation}
Q_{BVJ}^T = (B-V)^T - {\frac{E_{BV}}{E_{VJ}}}^T\,(V-J)^T 
           =(B-V)^S- {\frac{E_{BV}}{E_{VJ}}}^S\,(V-J)^S. 
\end{equation}
The difference between synthetic $Q_{BVJ}^T = (B-V)^T - {(E_{BV}/E_{VJ})}^T\,(V-J)^T$ 
and the ``observed'' index 
$Q_{BVJ}^S = (B-V)^S - {(E_{BV}/E_{VJ})}^S\,(V-J)^S$ in this ideal case will be zero. 
However, we know nothing about parameters 
of the star except our assumption that the stellar parameters are inside our synthetic grid. 

Let us try each of the grid points for the input colors of the star. We take the difference 
\begin{equation}
\Delta Q_{BVJ} =|Q_{BVJ}^T-(B-V)^S + {\frac{E_{BV}}{E_{VJ}}}^T\,(V-J)^S|
\end{equation}
in each grid point. We do the same for $\Delta Q_{VJH}$ and $\Delta Q_{JHK_S}$. 
All points that satisfy the 
following equations

\begin{eqnarray}
\label{sel_crit}
\Delta Q_{BVJ}   & < \sigma_{Q_{BVJ}} \nonumber    \\
\Delta Q_{VJH}   & < \sigma_{Q_{VJH}}              \\
\Delta Q_{JHK_S} & < \sigma_{Q_{JHK_S}} \nonumber
\end{eqnarray}
are selected as possible solutions. 

The definition of $\sigma_{Q_{BVJ}}$ comes from the definition of Q: 
\begin{equation}
\sigma_{Q_{BVJ}}^2
=\sigma_{(B-V)^S}^2+({\frac{E_{BV}}{E_{VJ}}}^T)^2\,\sigma_{(V-J)^S}^2+((V-J)^S)^2\,\sigma_{{E_{BV}/E_{VJ}}^T}^2.
\end{equation}
The problem are the quantities of $(E_{BV}/E_{VJ})^T$ and $\sigma_{{E_{BV}/E_{VJ}}^T}^2$, 
which are unknown a priori. 
In this step we take the 
maximum value of $(E_{BV}/E_{VJ})^T$ in our theoretical grid (averaged over all points in the grid 
$<(E_{BV}/E_{VJ})^T> = 0.426466 \pm 0.156561$, maximum value 
$ \textrm{max}\,(E_{BV}/E_{VJ})^T = 0.4516$) for all points, 
and double the value of $({E_{BV}/E_{VJ}}^T)^2\,\sigma_{(V-J)^S}^2$ as a rough estimation of 
${(V-J)^S}^2\,\sigma_{{E_{BV}/E_{VJ}}^T}^2$. 
Finally, 
\begin{equation}
\sigma_{Q_{BVJ}}^2 =\sigma_{(B-V)_S}^2+0.9032\,\sigma_{(V-J)_S}^2.
\end{equation}
The same is done for the estimation of $\sigma_{Q_{VJH}}$ and $\sigma_{Q_{JHK}}$.

After the initial selection of grid points in accordance with Eq.~(\ref{sel_crit}) 
we return from the extinction-free space 
to the color space and remove all ``artifact'' entries (those with negative extinction 
and colors out of the bounding box for the input star).

We assume that the ratio $A_B/A_V$ is constant for a given temperature, gravity, and metallicity 
and that is only depends on these 
parameters and not on  $A_V$ itself. 
  
For each of the selected grid points, we can find reddening for each color and 
extinction for each band, 
\begin{equation}
{E_{c}}^S =c^S-c^T  
\end{equation}
where $c^S$ is the observed color for the star and $c^T$ is the synthetic color at the selected
grid point, $c=(B-V),(V-J),(J-H),(H-K_S)$. Also,
\begin{equation}
A_V^{c} = (\frac{A_V}{E_{c}})^T\,{E_{c)}}^S,  
\end{equation}
where $(A_V/E_{c})^T$ is the synthetic ratio of extinction in the Johnson V band to color
excess in $c$ color at each grid point.

We use the inverse of the variances of the input colors as weights to find the weighted average  
for the extinction. 
From the previous equations and the definition of ${E_{c}}^S$ the weights are   
\begin{equation}
w_{c} = \frac{(E_{c}/A_V)_T^2}{\sigma_{c}^2}. 
\end{equation}
To remove negative $A_V$, only grid points with $A_V > -3\,\sigma_{A_V}$ were selected, 
where $\sigma_{A_V}$ is a weighted 
dispersion of $A_V$.

For each point in the grid we have theoretical magnitudes, colors, theoretical excess ratios, 
and an estimation of the extinction $A_V$. Now we calculate 
for each point a synthetic reddened color,
\begin{equation}
\label{new_colors}
c^T_{r} = c^T + (\frac{E_{c}}{A_V})^T\,A_V,  
\end{equation}
and remove points that are out of the $3\,\sigma$ range from the input color of the star. 
Finally, we select points with 
\begin{equation}
|c^T_{r} - c^S | < 3\,\sigma_{c^S} 
\end{equation}
for each color $c$ and have N grid points as possible solutions for the star.

\subsection{The second part of the method: cluster analysis}

We selected N points that satisfy our requirement (Eq.~\ref{sel_crit}), that are within the
allowed extinction range ($A_V > -3\,\sigma_{A_V}$), and that are inside the error box of
the stellar colors. All these points are possible solutions for the input star  
must be grouped properly to form a set of possible solutions in the next step. 

The distribution of the grid points is uniform in the parameter space 
($\log T_{eff}, \log g$, [M/H]). We selected all grid points that are inside 
the error box of
stellar colors, so are now able to simulate a more realistic physical distribution of grid points
inside the error box assuming a normal distribution of errors of the colors of the star. 
This step will
be made by giving new weights to each selected gridpoint.

The weighted distribution of grid points inside the error box will be projected onto the axes in the 
parameter space to subdivide the subspace with selected grid points into the set of possible 
solutions and to assign a probability to each solution.

\subsubsection{Weighting of points}
\label{weight-points}

\begin{figure}
\centering
\includegraphics[width=9cm]{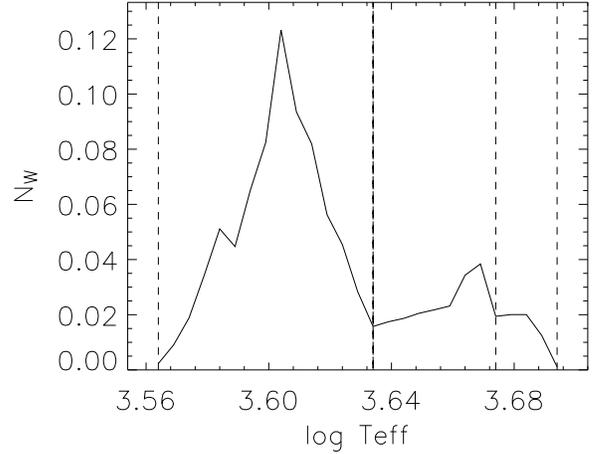}
   \caption{ The projection of the weighted distribution function of selected grid points on the 
   $\log T_{eff}$ axis, showing the subdivision made with an assumed 10\% significance level.
 	   }
      \label{Fig_distr_teff}
\end{figure}

\begin{figure}
\centering
\includegraphics[angle=270,width=9cm]{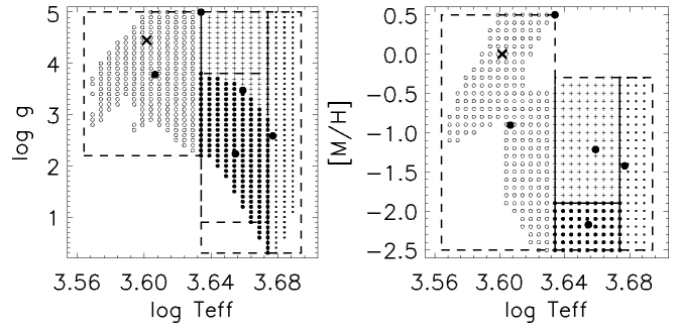}
   \caption{Points selected for the test input point in the parameter space. 
   Big open circles are for 
   solution 1 (see Table~\ref{set_sol}), big closed circles for solution 2, crosses 
   for solution 3, and small circles are for solution 5.  
 	   }
      \label{Fig_param_diag}
\end{figure}

\begin{table*}
\caption{Set of solutions for a test point of $T_{eff}=4000$ K, $\log g=4.45$ and [M/H]=0.0 and 
with an assumed error in color of 0.05 mag.}
\label{set_sol}
\begin{tabular}{|c|c|c|c|c|c|c|c|c|c|c|}
\hline
Number    & Total  & $T_{eff}^{min}$, & $T_{eff}^{max}$, &  $T_{eff}^w \pm \sigma_{T_{eff}}$, & 
$\log g^{min}$, & $\log g^{max}$, &  $\log g^w \pm \sigma_{\log g}$,& $[M/H]^{min}$ & $[M/H]^{max}$
& $[M/H]^{w} \pm \sigma_{[M/H]}$  \\ 
of points & weight &     K            &       K          &      K        & & & & & & \\
\hline
1713 & 0.7204 & 3664 & 4305 & $4041^{-147}_{+153}$ &  2.2 & 5.0 & 3.8 $\pm$ 0.4 & -2.5  &  0.5 &  -0.9 $\pm$  0.8  \\
951  & 0.0744 & 4305 & 4721 & $4515^{-151}_{+156}$ &  0.3 & 3.8 & 2.2 $\pm$ 0.5 & -2.5  & -1.9 &  -2.2 $\pm$  0.04\\
2736 & 0.1346 & 4305 & 4721 & $4561^{-145}_{+150}$&  0.9 & 5.0 & 3.5 $\pm$ 0.7 & -1.9  & -0.3 &  -1.2 $\pm$  0.2\\
1    & 0.0006 & 4305 & 4305 & $4305$ &  5.0 & 5.0 & 5.0           &  0.5  &  0.5 &   0.5  \\
1728 & 0.0700 & 4721 & 4943 & $4754^{-51}_{+51}$&  0.3 & 5.0 & 2.6 $\pm$ 1.3 & -2.5  & -0.3 &  -1.4 $\pm$  0.4\\
\hline
\end{tabular}	   
\end{table*}

\begin{figure}
\centering
\includegraphics[width=9cm]{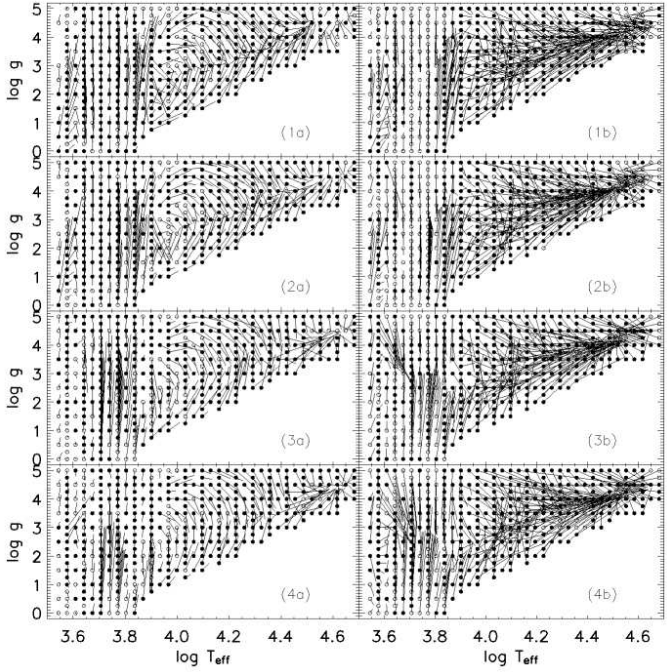}
   \caption{The offset (bias) in the case of the selection of the solution with 
   the largest weight in the 
   ($\log T_{eff}$,$\log g$) plane. Open
   circles are the cases of a single solution, lines show the distance between the selected
   solution and the input grid point. Panel (1a) is for metallicity $[M/H]=-2.0$ dex and 
   an assumed rms-error $\sigma_{ph}=0.01$
   mag in all colors, panel (1a) for metallicity $[M/H]=-2.0$ dex and 
   an assumed rms-error $\sigma_{ph}=0.05$ mag in all colors, 
   panel (2a) --- $[M/H]=-1.0$ dex and $\sigma_{ph}=0.01$ mag, 
   panel (2b) --- $[M/H]=-1.0$ dex and $\sigma_{ph}=0.05$ mag, 
   panel (3a) --- $[M/H]=0.0$ dex and $\sigma_{ph}=0.01$ mag, 
   panel (3b) --- $[M/H]=0.0$ dex and $\sigma_{ph}=0.05$ mag, 
   panel (4a) --- $[M/H]=0.3$ dex and $\sigma_{ph}=0.01$ mag, 
   panel (4b) --- $[M/H]=0.3$ dex and $\sigma_{ph}=0.05$ mag. 
 	   }
      \label{Fig_bias}
\end{figure}

\begin{figure}
\centering
\includegraphics[width=9cm]{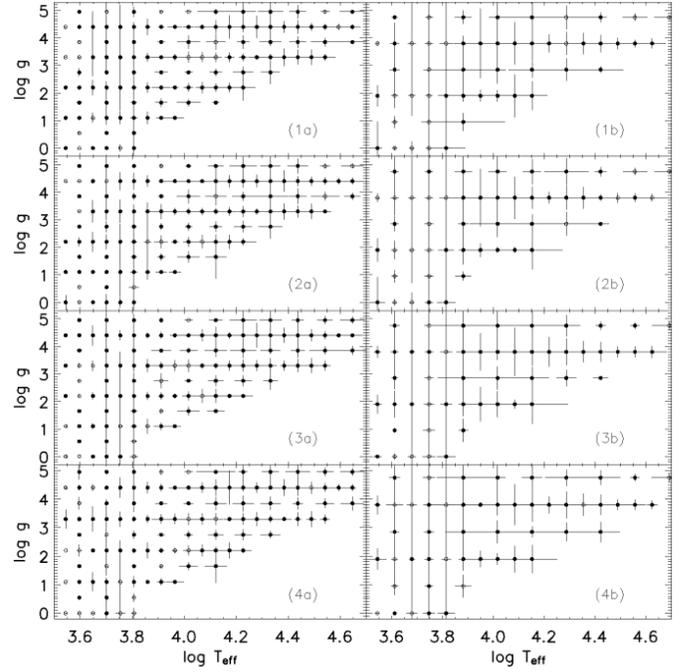}
   \caption{Errors of the selected solution on the ($\log T_{eff}$,$\log g$) plane.
     Panel (1a) is for metallicity $[M/H]=-2.0$ dex and 
   an assumed error $\sigma_{ph}=0.01$
   mag in all colors, panel (1a) for metallicity $[M/H]=-2.0$ dex and 
   an assumed error $\sigma_{ph}=0.05$ mag in all colors, 
   panel (2a) --- $[M/H]=-1.0$ dex and $\sigma_{ph}=0.01$ mag, 
   panel (2b) --- $[M/H]=-1.0$ dex and $\sigma_{ph}=0.05$ mag, 
   panel (3a) --- $[M/H]=0.0$ dex and $\sigma_{ph}=0.01$ mag, 
   panel (3b) --- $[M/H]=0.0$ dex and $\sigma_{ph}=0.05$ mag, 
   panel (4a) --- $[M/H]=0.3$ dex and $\sigma_{ph}=0.01$ mag, 
   panel (4b) --- $[M/H]=0.3$ dex and $\sigma_{ph}=0.05$ mag.  
 	   }
      \label{Fig_errors}
\end{figure}

\begin{figure}
\centering
\includegraphics[angle=270,width=9cm]{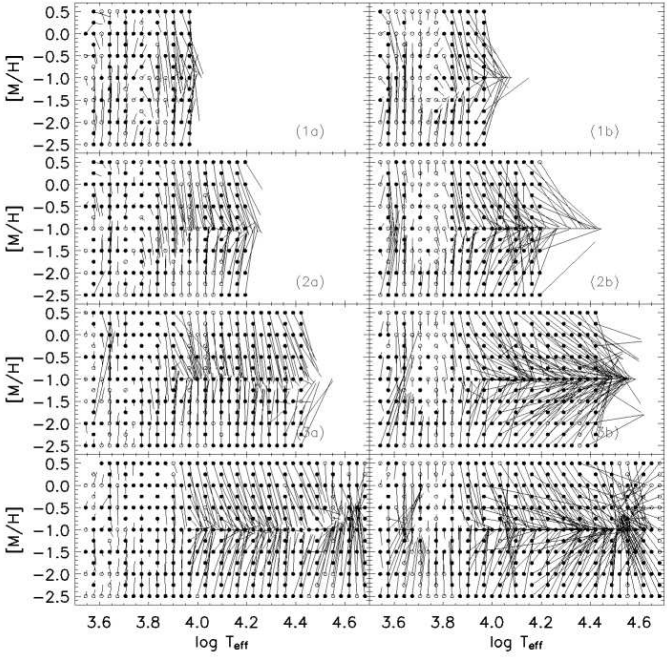}
   \caption{The bias in the case of the selection of the solution with the largest weight in the 
   ($\log T_{eff}$,[M/H]) plane. Open
   circles are the cases of a single solution, lines show the distance between the selected
   solution and the input grid point. Panel (1a) is for  $\log g=1.0$ dex and 
   an assumed error $\sigma_{ph}=0.01$
   mag in all colors, panel (1a) for $\log g=1.0$ dex and 
   an assumed error $\sigma_{ph}=0.05$ mag in all colors, 
   panel (2a) --- $\log g=2.0$ dex and $\sigma_{ph}=0.01$ mag, 
   panel (2b) --- $\log g=2.0$ dex and $\sigma_{ph}=0.05$ mag, 
   panel (3a) --- $\log g=3.0$ dex and $\sigma_{ph}=0.01$ mag, 
   panel (3b) --- $\log g=3.0$ dex and $\sigma_{ph}=0.05$ mag, 
   panel (4a) --- $\log g=4.5$ dex and $\sigma_{ph}=0.01$ mag, 
   panel (4b) --- $\log g=4.5$ dex and $\sigma_{ph}=0.05$ mag.  
 	   }
      \label{Fig_bias_met}
\end{figure}

\begin{figure}
\centering
\includegraphics[angle=270,width=9cm]{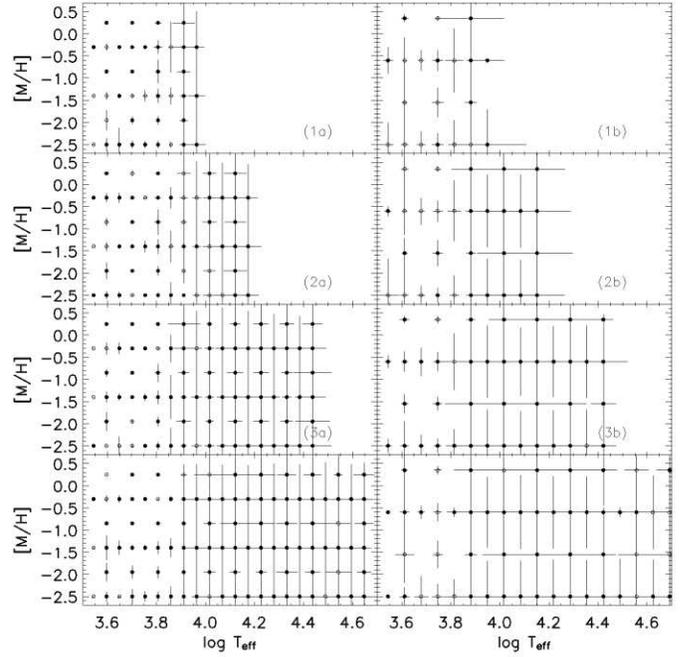}
   \caption{Errors in the case of the selection of the solution with the biggest weight on the 
   ($\log T_{eff}$,[M/H]) plane.  Panel (1a) is for  $\log g=1.0$ dex and 
   an assumed error $\sigma_{ph}=0.01$
   mag in all colors, panel (1a) is for $\log g=1.0$ dex and 
   an assumed error $\sigma_{ph}=0.05$ mag in all colors, 
   panel (2a) --- $\log g=2.0$ dex and $\sigma_{ph}=0.01$ mag, 
   panel (2b) --- $\log g=2.0$ dex and $\sigma_{ph}=0.05$ mag, 
   panel (3a) --- $\log g=3.0$ dex and $\sigma_{ph}=0.01$ mag, 
   panel (3b) --- $\log g=3.0$ dex and $\sigma_{ph}=0.05$ mag, 
   panel (4a) --- $\log g=4.5$ dex and $\sigma_{ph}=0.01$ mag, 
   panel (4b) --- $\log g=4.5$ dex and $\sigma_{ph}=0.05$ mag.  
 	   }
      \label{Fig_errors_met}
\end{figure}

\begin{figure}
\centering
\includegraphics[width=9cm]{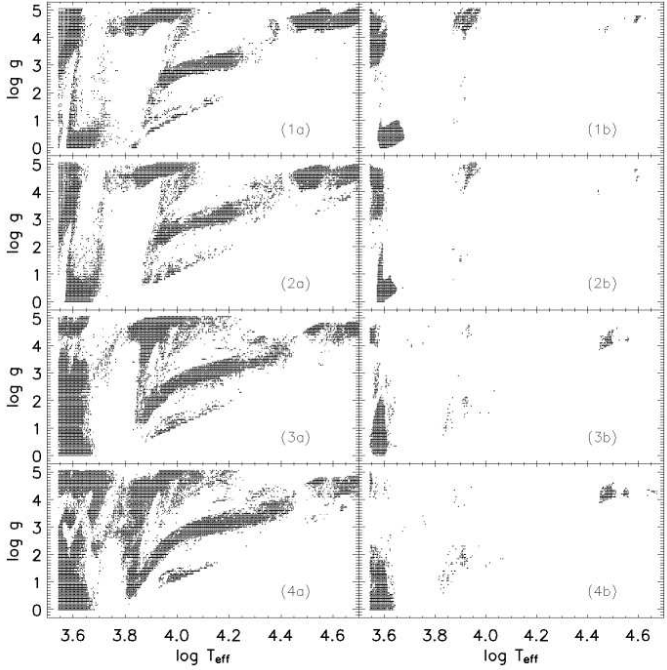}
   \caption{The best regions for estimating astrophysical parameters 
    on the 
   ($\log T_{eff}$,$\log g$) plane. Biases and errors of the solution (see text 
   for the explanation) are in the range: $\Delta \log T_{eff} <  0.05$, 
   $\sigma_{\log T_{eff}} < 0.05$, 
   $\Delta \log g < 0.3$, $\sigma_{\log g} < 0.3$. 
   Panel (1a) is for metallicity $[M/H]=-2.0$ dex and 
   an assumed error $\sigma_{ph}=0.01$
   mag in all colors, panel (1a) for metallicity $[M/H]=-2.0$ dex and 
   an assumed error $\sigma_{ph}=0.05$ mag in all colors, 
   panel (2a) --- $[M/H]=-1.0$ dex and $\sigma_{ph}=0.01$ mag, 
   panel (2b) --- $[M/H]=-1.0$ dex and $\sigma_{ph}=0.05$ mag, 
   panel (3a) --- $[M/H]=0.0$ dex and $\sigma_{ph}=0.01$ mag, 
   panel (3b) --- $[M/H]=0.0$ dex and $\sigma_{ph}=0.05$ mag, 
   panel (4a) --- $[M/H]=0.3$ dex and $\sigma_{ph}=0.01$ mag, 
   panel (4b) --- $[M/H]=0.3$ dex and $\sigma_{ph}=0.05$ mag.  
 	   }
      \label{Fig_region}
\end{figure}

\begin{figure}
\centering
\includegraphics[angle=270,width=9cm]{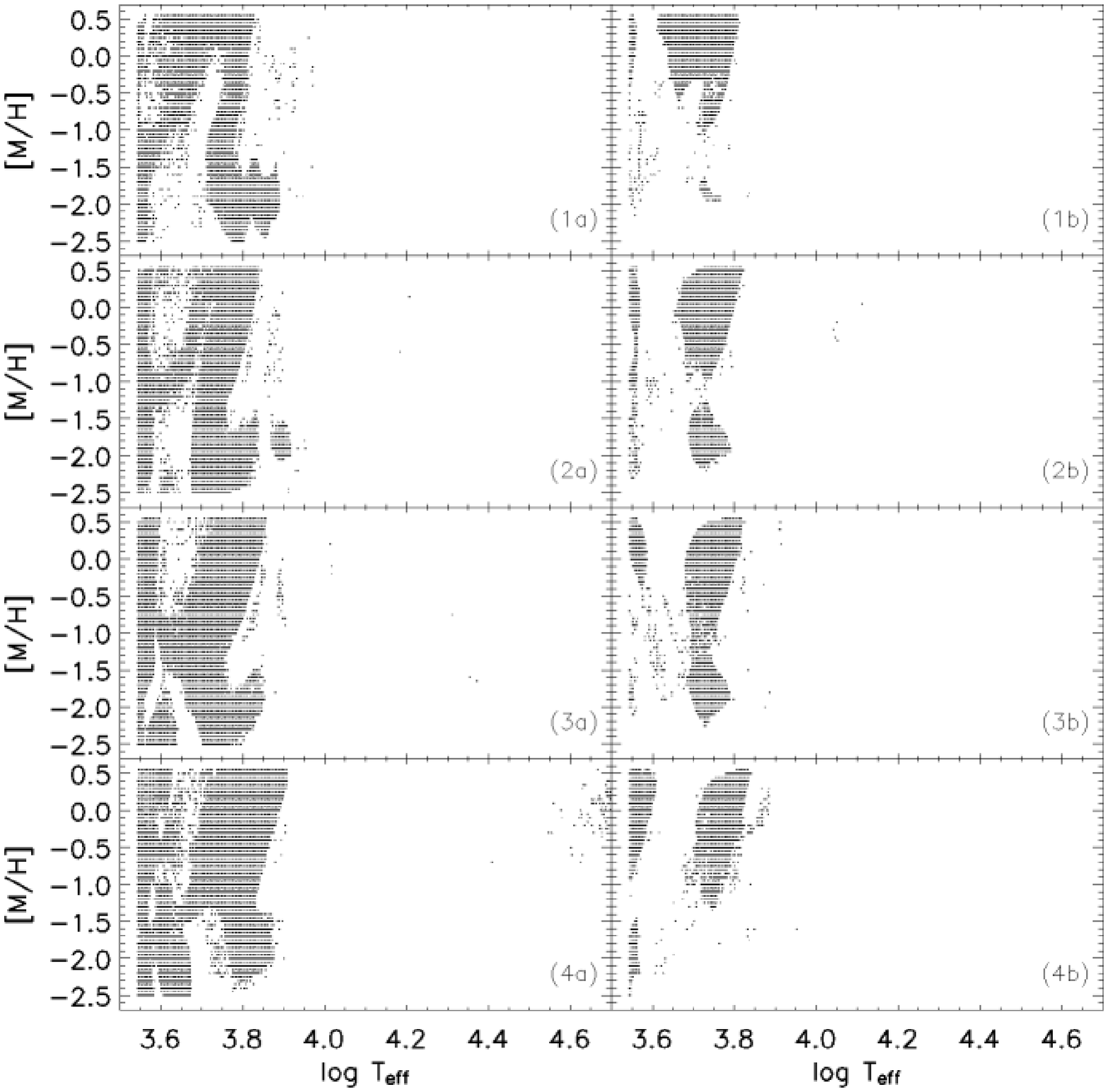}
   \caption{The best regions for estimating astrophysical parameters 
    on the 
   ($\log T_{eff}$,[M/H]) plane. Biases and errors of the solution (see text 
   for the explanation) are in the range: $\Delta \log T_{eff} <  0.05$, 
   $\sigma_{\log T_{eff}} < 0.05$, 
   $\Delta [M/H] < 0.3$, $\sigma_{[M/H]} < 0.3$.Panel (1a) is for  $\log g=1.0$ dex and 
   an assumed error $\sigma_{ph}=0.01$
   mag in all colors, panel (1a) for $\log g=1.0$ dex and 
   an assumed  error $\sigma_{ph}=0.05$ mag in all colors, 
   panel (2a) --- $\log g=2.0$ dex and $\sigma_{ph}=0.01$ mag, 
   panel (2b) --- $\log g=2.0$ dex and $\sigma_{ph}=0.05$ mag, 
   panel (3a) --- $\log g=3.0$ dex and $\sigma_{ph}=0.01$ mag, 
   panel (3b) --- $\log g=3.0$ dex and $\sigma_{ph}=0.05$ mag, 
   panel (4a) --- $\log g=4.5$ dex and $\sigma_{ph}=0.01$ mag, 
   panel (4b) --- $\log g=4.5$ dex and $\sigma_{ph}=0.05$ mag.   
 	   }
      \label{Fig_region_met}
\end{figure}

Suppose that the ``observed'' star with colors $(B-V)^S, (V-J)^S, (J-H)^S, (H-K_S)^S$ has a normal
distribution of errors in each color ($N(0,\sigma_c)$) and all colors are independent variables.
From Eq.~(\ref{new_colors}) we can find the distance of each grid point from the center of 
the error box (from the observed color of star), 
\begin{equation}
\Delta c = c^T_{r} - c^S, 
\end{equation}
where c is one of the colors, $c=(B-V),(V-J),(J-H),(H-K_S)$.

The probability that a point occupies the 4-dimensional 
color space from $(B-V)$ to $(B-V)+d(B-V)$, $(V-J)$ to $(V-J)+d(V-J)$, $(J-H)+d(J-H)$,
$(H-K_S)+d(H-K_S)$ is 
\begin{eqnarray}
dN = \prod_{\textrm{c}=(B-V),\,(V-J),\,(J-H),\,(H-K_S)} f(\Delta \textrm{c})\,d\textrm{c},
\end{eqnarray}
where 
\begin{equation}
f(\Delta (B-V))=e^{-\frac{(\Delta (B-V))^2}{\sigma_{B-V}^2}}.
\end{equation}
The same distribution for other colors is assumed. 

The real distribution of grid points is $N(\Delta (B-V), \Delta (V-J), \Delta (J-H), \Delta (H-K))$. 
This is a number of points with $\Delta (B-V),\,\Delta(V-J),\,\Delta (J-H),\,\Delta (H-K)$ between $ \Delta (B-V)$
and $\Delta (B-V) + d(B-V)$, $\Delta (V-J)$ and $\Delta (V-J) + d(V-J)$, and so on; i.e. 
$N_p =N(\Delta (B-V),\Delta (V-J),\Delta (J-H),\Delta (H-K))\,d(B-V)\,d(V-J)\,d(J-H)\,d(H-K) $. 

The one-dimensional weight for the color (B-V), for example, is 
$w_{\Delta (B-V)}=f(\Delta (B-V))\,d(B-V)$ and for a point 
in the 4-dimensional color space 
\begin{eqnarray}
w^i=\frac{\prod\limits_{c=(B-V),(V-J),(J-H),(H-K_S)} w_{\Delta c}}{N_p}.
\end{eqnarray}
The optimal bin in the case of the one-dimensional distribution is 
$h = 3.5\,\sigma\,n^{-1/3} $
and, in the case of d dimensions, the optimal bin is $h=O(n^{-1/(d+2)})$ 
(see Scott et al.~\cite{Scott}). 
After the weighting of the points we normalise the weight of all selected points to 1, so that 
$ \sum_i w^i =1 $.

\subsubsection{The final set of solutions}

The distribution of the weighted grid points was projected on each axis in the parameter space 
($\log T_{eff}$, $\log g$, [M/H]) to group the grid points into clusters in the parameter space. 
These clusters are the final set of possible solutions for the input star. 

Let us suppose that the grid points are distributed between a minimum  
$\log T_{eff}^{min}$ and a maximum  $\log T_{eff}^{max}$. We used 
the original bin of the grid ($\Delta \log T_{eff}=0.005$ dex) to construct 
the projection of the distribution of N selected grid points on the temperature axis:   
\begin{equation}
f_{\log T_{eff}^i} = \sum_{j=1}^{N} w_j,  w_j : \log T_{eff}^j \equiv \log T_{eff}^i,   
\end{equation}
where i runs through 1 to $M$, and $M$ is the number of bins of the grid between $\log T_{eff}^{min}$
and $\log T_{eff}^{max}$, $M=(\log T_{eff}^{max} - \log T_{eff}^{min})/ \Delta \log T_{eff}$. 
Note that $\log T_{eff}$, $\log g$ and [M/H] are discrete values of the grid. 

Let us suppose that we have found $L$ local maxima in $\log T_{eff}$ space. 
We estimate the significance of each local
maximum in accordance with the maximum likelihood criterion for the significance of the 
maximum (see Materne~\cite{materne}), i.e. for the k-th local maximum the criteria for significance 
is 
\begin{equation}
\chi_k^2 = -2\,ln \frac{L_{-k}}{L},
\end{equation}
where 
\begin{equation}
L = \Pi_{j=1}^M f_{\log T_{eff}^j},
\end{equation}
and $L_{-k}$ is the likelihood in the absence of the k-th local maximum.

We use the 10\% significance level ($\chi^2 > 2.7$). If $\chi_k \leq 2.7 $ the maximum 
is neglected and joined with the neighboring, most significant maximum.  
This procedure decreases the number of clusters in $\log T_{eff}$ space to $N \leq L$.
In each of
these clusters, we are looking for a clustering on the $\log g$ axis, and for each cluster on the $\log g$
axis, we are looking for the clustering on the metallicity axis with the same procedure. 
For example, 
there are $N_{\log T_{eff}}$ clusters on the temperature axis, 
the i-th cluster on the temperature 
axis has $N_{\log g}^i$ clusters on the gravity axis and the j-th cluster on the gravity axis has 
$N_{[M/H]}^{i,j}$ clusters on the metallicity axis. 

The final number of clusters (and this is the final set of solutions) is 
\begin{equation}
N_{clusters} = \sum_{i=1}^{N_{\log T_{eff}}}\,\sum_{j=1}^{N_{\log g}^i}\,N_{[M/H]}^{i,j}. 
\end{equation}
We have distributed N grid points selected over $N_{clusters}$ clusters, 
the sum of weights of points 
inside the cluster gives us the weight of each cluster: 
\begin{equation}
W_j = \sum w_i, w_i :i \in N_j,  
\end{equation}
where $N_j$ is the j-th cluster. 

To find average values 
of temperature, gravity, and metallicity for each cluster, we repeated the procedure of 
weighting described in Sect.~\ref{weight-points}, this time inside each cluster, and 
compute the weighted average solutions for $\log T_{eff}$, $\log g$, [M/H], $A_V$, and 
distance r with the dispersion 
of these values inside the cluster.

\subsection{The third part of the method: selection of the solution}

We forme a set covering all possible solutions for a star. The final and
most complicated step is the selection of one solution from the set.

As we see from Fig.~\ref{Fig_2c} there is no possibility of selecting 
a single ``true'' solution
in general. Nevertheless we can select ``the statistically most probable'' solution 
by the appropriate criteria. 
We would like to underline that, by these criteria, we do not use guesses about the astrophysical parameters 
themselves, so the selection at this point does not introduce limitations on the parameters. 

There are a number of methods for this selection: we can take the solution with 
the largest weight or
with minimal distance from the input ``observed'' color in color space or take one 
with an unbiased distribution
of grid points around the center of the error box in color space.

Let us show an example of the determination of parameters to point out the major troubles we have to face. 
We used the point from the synthetic grid with $T_{eff}=4000$ K, $\log g=4.45$, and [M/H]=0.0, 
photometric errors equal to 0.05 mag in all colors were taken. The first step of the method returned 6000
points distributed between 3664 K and 4943 K in temperature, between 0.3 and 5.0 dex in $\log g$
and between -2.5 and 0.5 dex in metallicity. 
All points were weighted and grouped in 5 possible 
solutions (see Table~\ref{set_sol}). 
The second part of the method (the cluster analysis) is illustrated 
by Fig.~\ref{Fig_distr_teff}. The first maximum on the $\log T_{eff}$ axis is rejected 
by the significance level assumed.
Finally Fig.~\ref{Fig_param_diag} presents the clusters in the projections onto the 
$\log T_{eff} - \log g$ and the $\log T_{eff} - [M/H]$ planes.

We define the following two quantities that describe the error of the estimated parameter: 
\begin{itemize}
\item[1.] The offset (bias) of the selected solution from the true parameters $\Delta \log T_{eff}$, 
$\Delta \log g$, $\Delta [M/H]$. These values are the differences between the estimated 
parameter (the center of the selected cluster) and the true parameter known only in simulations. 
\item[2.] The dispersion of the estimated parameter  
$\sigma_{\log T_{eff}}$, $\sigma_{\log g}$, $\sigma_{[M/H]}$.  
\end{itemize}

From Fig.~\ref{Fig_param_diag} we see that it is possible to subdivide the first cluster (first 
solution in Table~\ref{set_sol}). This possibility was ruled out by the significance level we
assumed (see Fig.~\ref{Fig_distr_teff}, the first peak on the distribution function). If we
decrease the significance level, we are increasing the number of clusters in the set of 
solutions  but 
decrease the weight for each cluster. Selecting a single cluster we face the problem of 
choosing between 
a number of clusters: the one that includes the true parameters of a star we call ``true''
cluster (solution) for the star, the remaining are ``false'' solutions. In practice 
we can distinguish 
between ``true'' and ``false'' solutions only if we knew the parameters of the star a priori; 
however, 
we can determine the region in the parameter space where only a single, hence ``true'', solution 
is available or where the ``true'' solution can be selected from a set of solutions by some
criteria.  We  note that 
the definition of this region in parameter space depends on the desired precision of the 
determination of parameters. We show below how to define such a ``most preferable'' region.

\subsubsection{The most preferable regions in parameter space}

\begin{figure*}
\centering
\includegraphics[angle=270,width=16cm]{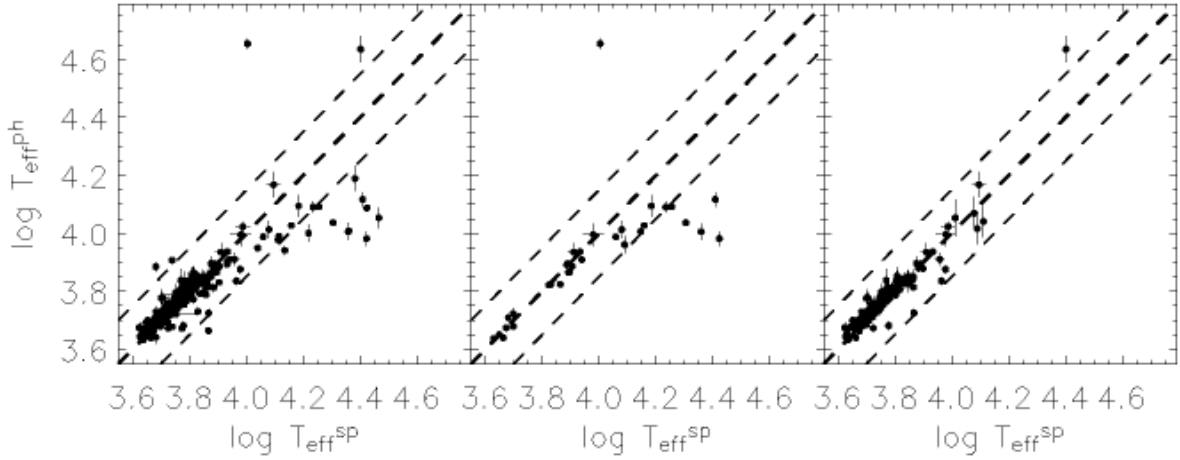}
   \caption{The test of the method on the calibration sample estimating temperature. 
   In the left panel there are all stars with $\sigma_{\log T_{eff}} < 0.05$ for 
   an estimated temperature 
   and errors of photometry 
   better than 0.05 mag in all colors. The middle panel 
   shows a subsample of the left panel falling into the most suitable region on 
   the $(\log T_{eff}, \log g)$ plane
   as described in the text and the right panel 
   gives a subsample of the left panel with $A_V[mag]/d[kpc] \in [0.1,2.0]$. 
   The dashed lines confine the 3-$\sigma$ 
   region; i.e, they are the median line and $\log T_{eff}^{sp} = \log T_{eff}^{ph} + 0.15$, 
   $\log T_{eff}^{sp} = \log T_{eff}^{ph} - 0.15$ lines. 
 	   }
      \label{Fig_logteff_comp_all}
\end{figure*}

\begin{figure*}
\centering
\includegraphics[angle=270,width=16cm]{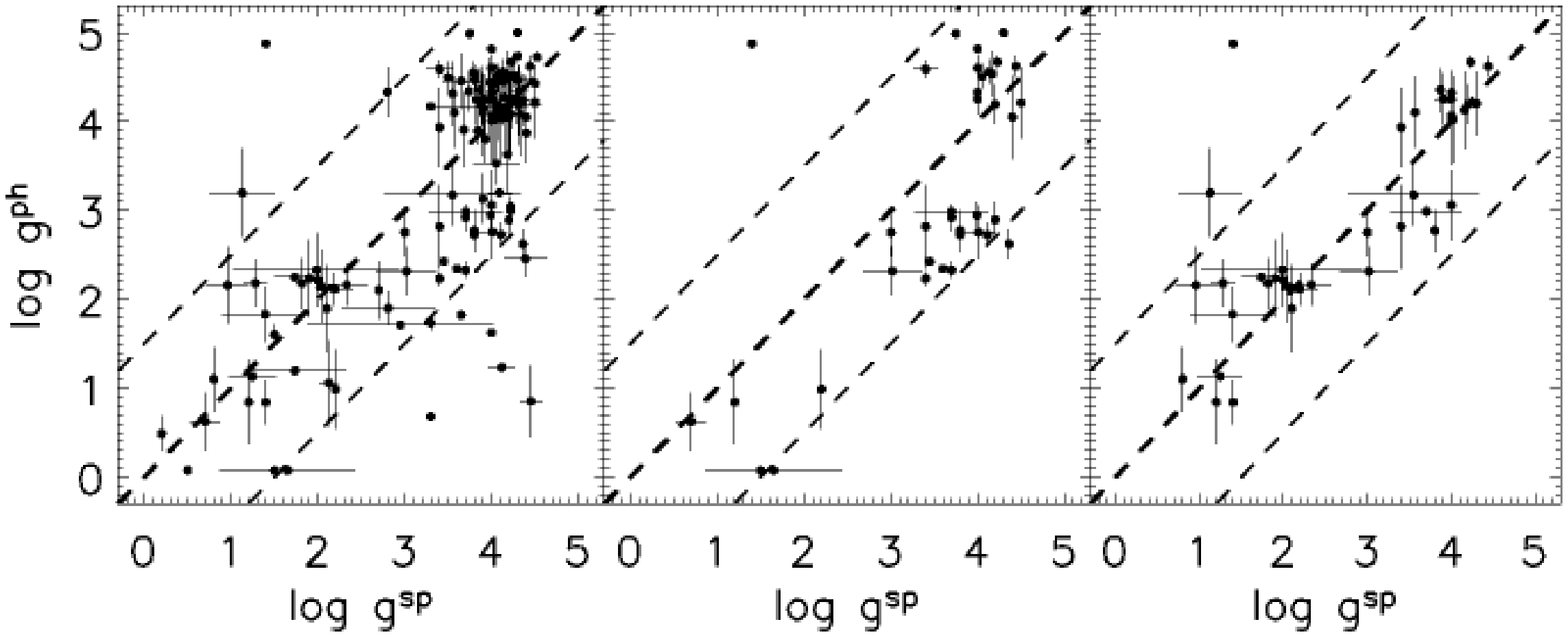}
   \caption{The test of the method on the calibration sample for estimating  gravity. 
   In the left panel there are all stars with $\sigma_{\log g} < 0.5$ for the 
   estimated gravity  
   and photometric errors  
   better than 0.05 mag in all colors. The middle panel 
   shows a subsample of the left panel falling into the most suitable region on 
   the $(\log T_{eff}, \log g)$ plane
   as described in the text, and  in the right panel, 
   we plot a subsample of the left panel with $A_V[mag]/d[kpc] \in [0.1,2.0]$. 
   The dashed lines are the median line and 
   $\log g^{sp} = \log g^{ph} + 1.5$, $\log g^{sp} = \log g^{ph} - 1.5$ lines.     
 	   }
      \label{Fig_logg_comp_all}
\end{figure*}

\begin{figure*}
\centering
\includegraphics[angle=270,width=16cm]{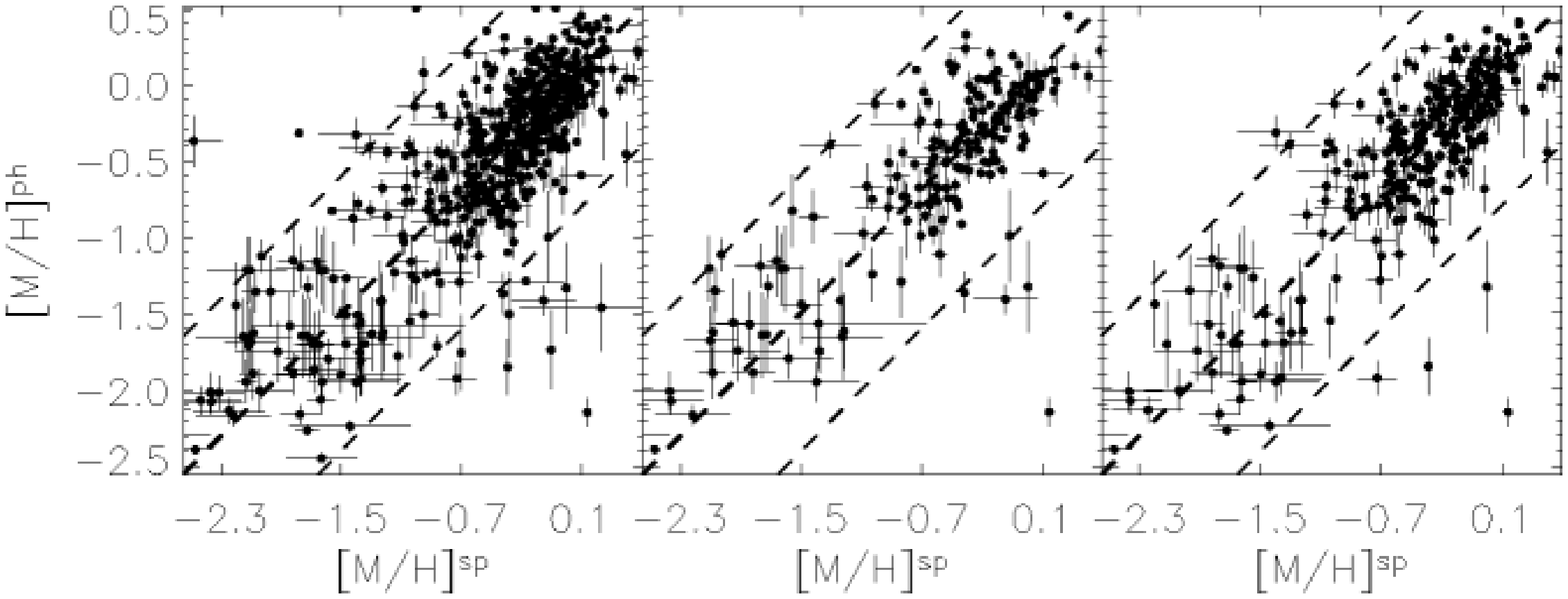}
   \caption{The test of the method on the calibration sample for estimating metallicity. 
   In the left panel there are all stars with $\sigma_{[M/H]} < 0.3$ for 
   the estimated metallicity and photometric errors  
   better than 0.05 mag in all colors. The middle panel 
   shows a subsample of the left panel falling into the most suitable region 
   on the $(\log T_{eff}, [M/H])$ plane
   as described in the text, and in the right panel, 
   we plot a subsample of the left panel with $A_V[mag]/d[kpc] \in [0.1,2.0]$. 
   The dashed lines are the median line and 
   $[M/H]^{sp} =[M/H]^{ph} + 0.9$, $[M/H]^{sp} = [M/H]^{ph} - 0.9$ lines. 
 	   }
      \label{Fig_logmet_comp_all}
\end{figure*}

\begin{table*}
\centering
\caption{Selection of the solution: comparison of different cases}
\label{stars_stat}
\begin{tabular}{|c|c|c|c|c|c|c|}
\hline
Parameter & \multicolumn{3}{|c|}{Number of stars} & \multicolumn{3}{|c|}{average and variance} \\
\cline{2-7}
          & criterion & criterion & criterion & criterion & criterion & criterion\\
	  &   1    & 1 \& 2 & 1 \& 3 &   1    & 1 \& 2 & 1 \& 3    \\
\hline
$\log T_{eff}^{sp} -\log T_{eff}^{ph}  $ & 588 & 31 & 330 & $0.001 \pm 0.059$ & $0.054 \pm 0.176$ & $-0.010 \pm 0.025$\\
$\log g^{sp} - \log g^{ph}$ & 133 & 38 &  40 & $-0.166 \pm 0.924$ & $-0.336 \pm 1.047$ & $0.191 \pm 0.778$\\
$[M/H]^{sp} - [M/H]^{ph}  $         & 527 & 199 & 305 & $0.011 \pm 0.457$ & $-0.029 \pm 0.459$ & $0.017 \pm 0.407$\\
\hline
\end{tabular}	   
\end{table*}

We have determined regions in the parameter space where we can best estimate astrophysical 
parameters; in other words, if the solution falls into this region, we can be sure that 
this estimation is a correct, unbiased solution for the star.

To find the most preferable regions, we performed simulations using the synthetic grid points as 
input ``observed'' colors. We interpolated between the grid points to create 
from a grid point with \{$\log T_{eff}$, $\log g$, [M/H]\} new points with 
\{ $\log T_{eff} + i\,\Delta \log T_{eff}/2$,  
$\log g + j\,\Delta \log g/2$, $[M/H] + k\,\Delta [M/H]/2 $ \}, $i,j,k = [0,1]$, and  
$\Delta \log T_{eff}$, $\Delta \log g$, and $\Delta [M/H]$ are intervals between points 
of the initial grid in
temperature, gravity, and metallicity. 
For each point of this enhanced grid, we calculated
synthetic colors. The resulting test grid includes points with the largest scatter from the points of the 
initial grid in the parameter space.

We tested the method on two sets of errors in colors; in the first case, 
all colors had an error 0.01
mag and in the second all colors had an error 0.05 mag. 
Figure~\ref{Fig_bias} shows the solution selected by the maximum weight with offsets (biases) caused by the selection of 
the ``false'' cluster or the shift inside the selected ``true'' cluster, 
and Fig.~\ref{Fig_bias_met} is the same for the temperature-metallicity plane. 
Figure~\ref{Fig_errors} plots the rms-errors of the selected 
solution for the temperature - gravity plane and Fig.~\ref{Fig_errors_met} - for the 
temperature-metallicity plane. 

We define a region in the parameter space as suitable for the unbiased determination 
of astrophysical parameters, if we 
can select a solution for the input point inside the region, and this final solution has biases 
for temperature $\Delta \log T_{eff} <  0.05$, 
for gravity $\Delta \log g < 0.3$, 
rms-errors  $\sigma_{\log T_{eff}} < 0.05$, $\sigma_{\log g} < 0.3$ 
in the temperature-gravity plane 
and $\Delta \log T_{eff} <  0.05$,  $\Delta [M/H] < 0.3$, 
rms-errors  $\sigma_{\log T_{eff}} < 0.05$, $\sigma_{[M/H]} < 0.3$ 
in the temperature-metallicity plane.
This is done in Figs.~\ref{Fig_region} and~\ref{Fig_region_met}. According to these figures
we define the following regions in the $\log T_{eff} -\log g$ plane
\begin{itemize}
\item[] $\log T_{eff} < 3.75$, $\log g > 4.0$;
\item[] $\log T_{eff} \in [3.395,3.7]$, $\log g < 1.0$; 
\item[] $\log T_{eff} \in [3.8,4.05]$, $\log g > 4.5$;
\item[] $\log T_{eff} \in [3.9,4.2]$, $\log g \in [2,3]$;
\item[] $\log T_{eff} > 4.4$, $\log g > 4.0$ 
\end{itemize}
as the most suitable for the determination of $\log T_{eff}$ and $\log g$, 
and on the $\log T_{eff} - [M/H]$ plane 
\begin{itemize}
\item[] $\log T_{eff} \in [3.7,3.8]$ 
\end{itemize}
as the most suitable in the search for metallicity. 
We note that the definition of these regions depends on the 
assumed errors of photometry, i.e. observations; 
and in the case of real stars with different observational errors in different bands, 
the definition of the most
suitable regions can vary.

\subsubsection{Setting bounds on extinction}
\label{criteria}

Estimating the most preferable regions for our method made in the previous section is 
based on the assumption that 
photometric errors are the same in all bands. In the case of real observations, we would 
have to 
recalculate the estimation of errors and biases at any point on the 
synthetic grid for each combination of errors of colors for the input stars. This is possible but consumes enormous computational 
resources and slows the estimation. 

For the actual application of the method to observations, we propose the following procedure: 
we estimate the extinction per photometric
distance. We assume a maximum extinction in the galactic plane to be 2.0 mag/kpc in 
the Johnson V band and a minimum extinction 
in the direction of the galactic poles as 0.1 mag/kpc (from Schlegel et al.~\cite{SFD}). 

Finally there are three criteria for selecting a single solution for the star: 
\begin{itemize}
\item[1.] selection of the solution with the largest weight,
\item[2.] selection based on the most preferable region in the parameter space, 
\item[3.] selection based on setting a bound  on the extinction per photometric distance.
\end{itemize}

\section{The method used in practice and problems of the method}

\begin{figure}
\centering
\includegraphics[angle=270,width=9cm]{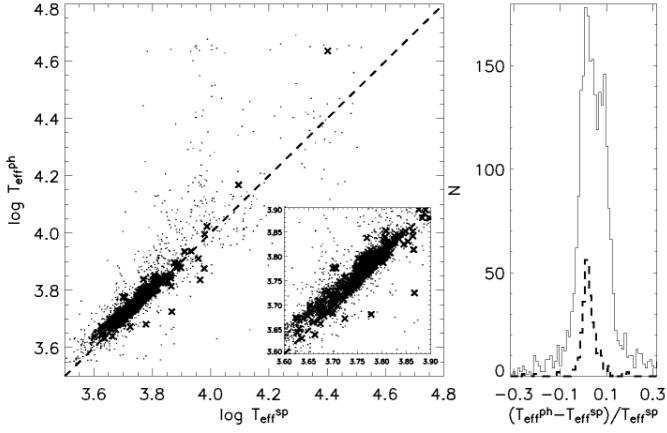}
   \caption{Comparison between the temperature determined from photometry (our method) 
   and spectroscopy (compiled catalog).
   Dots are all stars independent of the weight of the solution and the rms-error of the 
   estimated
   temperature, crosses are stars with a weight of the solution $w > 0.8$, photometric 
   errors better 
   than 0.05 mag in all bands, $\sigma_{\log T_{eff}^{ph}} < 0.05$ and $A_V[mag]/d[kpc] \in [0.1,2.0]$
   (criterion 1 \& 3 as described in Sect.~\ref{criteria}). 
   The right panel shows the corresponding
   histograms for both subsamples.}
      \label{Fig_teff_comp_fin}
\end{figure}

We used the 2096 stars of our compiled catalog (see Sect.~\ref{calibration} for details)
 with Johnson $B$, $V$, and 2MASS $J$, $H$, $K_S$ photometry to test the method in practice.

Table~\ref{stars_stat} shows the statistics for the different combinations of the criteria described in the previous 
section: criterion 1,
criterion 1 combined with criterion 2 for selected stars 
(we select a single solution with the largest weight for each star and 
select stars within the most preferable regions), and criterion 1 combined with criterion 3 
for selected stars.  
The last 3 columns in the table show the means and variances 
of the differences between the ``true'' spectroscopic parameter and our estimation (for example, 
$<\log T_{eff}^{sp}-\log T_{eff}^{ph}> \pm \sigma_{\log T_{eff}^{sp}-\log T_{eff}^{ph}}$). 
The results for the calibration subset, based on the different selection criteria are shown in
Figs.~\ref{Fig_logteff_comp_all},~\ref{Fig_logg_comp_all},~\ref{Fig_logmet_comp_all} 
(left panel for criterion 1, middle panel for criterion 1 \& 2, right panel for criterion 1 \& 3). 

\begin{figure}
\centering
\includegraphics[angle=270,width=9cm]{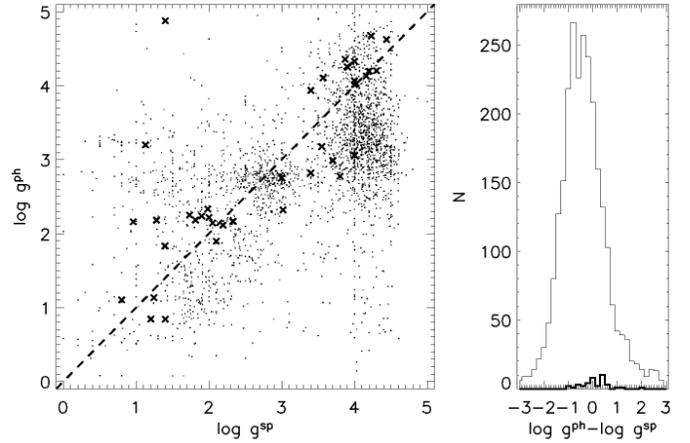}
   \caption{Comparison between gravities determined from photometry (our method) and spectroscopy
   (compiled catalog).
   Dots are all stars independent of the weight of the solution and the rms-error of the estimated
   gravity, crosses are stars with a weight of the solution $w > 0.8$, photometric errors 
   better 
   than 0.05 mag in all bands, $\sigma_{\log g^{ph}} < 0.5$ and $A_V[mag]/d[kpc] \in [0.1,2.0]$
   (criterion 1 \& 3 as described in Sect.~\ref{criteria}).
   The right panel shows the corresponding histograms for both subsamples.}
      \label{Fig_logg_comp_fin}
\end{figure}

\begin{figure}
\centering
\includegraphics[angle=270,width=9cm]{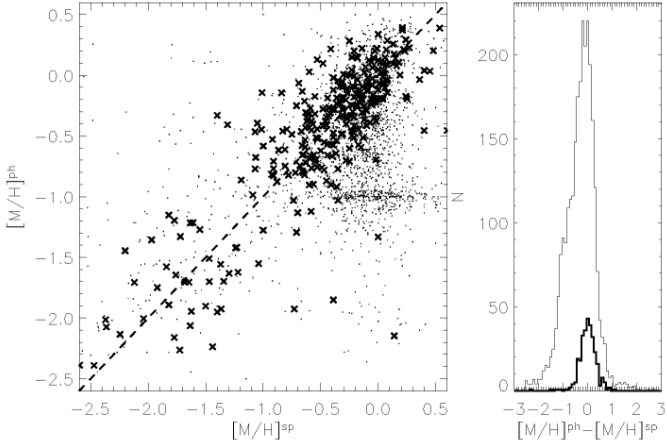}
   \caption{Comparison between metallicities determined from photometry (our method) and spectroscopy (compiled catalog).
   Dots are all stars independent of the weight of the solution and the rms-error 
   of the estimated
   metallicity, crosses are stars with a weight of the solution $w > 0.8$, 
   photometric errors better 
   than 0.05 mag in all bands, $\sigma_{[M/H]} < 0.3$ and $A_V[mag]/d[kpc] \in [0.1,2.0]$
   (criterion 1 \& 3 as described in Sect.~\ref{criteria}). 
   The right panel shows the corresponding histograms for both subsamples.}
      \label{Fig_logmet_comp_fin}
\end{figure}

The results for all stars in the compiled catalogue are shown in 
Figures~\ref{Fig_teff_comp_fin},~\ref{Fig_logg_comp_fin}, and~\ref{Fig_logmet_comp_fin}. 
We selected stars that have solutions with weight $w_n > 0.8$, 
uncertainties  of the input photometry better that 0.05 mag, rms-errors of the 
estimated parameters (dispersion of the points around the weighted average parameter for the solution) better that 0.05 dex 
for $\log T_{eff}$, 0.5 dex for $\log g$, and 0.3 dex for [M/H]. 
It becomes clear that gravity is the least accurate
unknown in  
estimating of astrophysical parameters: in the case of calibration catalog we were able to estimate gravity only 
for 2 \% of the stars with an appropriate precision (better than 0.5 dex). We note that only 1186 stars from the 
calibration catalog have photometric errors better than 0.05 mag and only 21 stars have photometric errors better than 0.02 mag 
in all bands.
We discuss the reasons for failures in the estimation of parameters below.   
We conclude that the combination of criteria 1 \& 3 works best in selecting stars 
with the best estimation of 
astrophysical parameters.

\subsection{The problem of the selection of the synthetic model grid}

There are a number of  libraries of synthetic models of atmospheres is available today. 
The most popular among them are the BaSeL library 
based on Kurucz models with an addition of synthetic and semi-empirical models for cool stars (Lejeune et al.,~\cite{BaSeL}), 
original Kurucz models (ATLAS9 program, Castelli \& Kurucz,~\cite{Kurucz2006}), 
and MARCS models (Gustafsson et al.,~\cite{MARCS}).
Even the models of the same author for the same parameters (microturbulent velocity etc.) 
differ due to improvements
that the authors introduced into different versions of models. 

We do not discuss the different choices of theoretical models leaving this subject to the user 
of the method, but we note that 
some of the discrepancies in the parameters estimated via our method in comparison with the parameters from spectroscopic estimations can 
be due to the different model sets used for photometric (performed by us) and spectroscopic (represented in the calibration catalog) 
estimations of astrophysical parameters. 

The direct comparison of Kurucz models with the SEDs of real stars (see, for example, Strai\v{z}ys et al.,~\cite{Straizys-Kurucz}) 
shows problems with the ultraviolet part of spectra for main-sequence stars, as well as 
an inconsistency in the synthetic SEDs with real 
SEDs for late type stars (K7-M stars). This will not influence our results for real stars as 
the lowest temperature estimation in 
Fig.~\ref{Fig_teff_comp_fin} turned out to be 4200 K but must be taken into consideration in future work as late type stars are in the 
area with the best estimation of parameters possible (see Figs.~\ref{Fig_region} and~\ref{Fig_region_met}).

Both photometric and spectroscopic determinations of astrophysical parameters based on
synthetic models suffer from the uncertainties and systematic errors of the selected synthetic grid.
We urge the users of this method to always make a reference to the synthetic models used and 
to accept the resulting parameters as parameters {\it in the system of the selected synthetic 
models}. A detailed study of the systematic errors 
of astrophysical parameters due to problems of synthetic models is beyond the scope of this paper.
There are many reasons for such systematics, for example, 
the convection treatment adopted for models and the influence of
models on fundamental parameters is discussed in Heiter et al.~(\cite{Heiter}), 
who estimate that the 
systematic effect in temperature could be up to 400 K.

The estimation of the systematics in the photometry due to the difference between synthetic models 
(Martins \& Coelho,~\cite{martins}) shows that, for Johnson B, V, J, H, and K
bands, they are within 0.05 mag, which can be accepted as an upper limit for this type 
of uncertainties in photometry.
The rough estimation of the influence of these uncertainties on the resulting parameters 
can be done in the same 
way as the estimation of biases due to errors in the input photometry ( 
Figs.~\ref{Fig_region} and~\ref{Fig_region_met}).

\subsection{The problem of stars outside the grid}

As we can see from the previous section and Figs. ~\ref{Fig_logteff_comp_all},~\ref{Fig_logg_comp_all},~\ref{Fig_logmet_comp_all} 
we reached quite good agreement between spectroscopic and photometric estimation of astrophysical parameters, however there are 
some cases of a large differences between our estimated parameters and parameters from spectroscopy.  
To understand the reason for these differences 
we searched the Simbad database 
for the information about each star with ``true'' (spectroscopic) parameter out of the 3-$\sigma$ range from the estimated parameters. 
Table~\ref{fail_reas} summarizes the results  and
indicates the reason for the difference between them.

The two stars with the greatest differences in the estimation of temperature are stars with peculiarities in the spectra. 
One of the cases where we failed to determine gravity is a Cepheid. 
For estimating the metallicity, we failed in the case of two close binaries, 
an RS CVn-like variable 
and a Cepheid (HD 101602, both gravity and metallicity were estimated wrong for this variable star). 

There is only one star for which the reason for the failure is unclear: BD~+30~2611. This star 
is a suspected Hipparcos binary star, although the radial velocity measurements did not 
prove the binarity (Sperauskas \& Bartkevicius~\cite{BD}).  

In all cases of failure, the stars 
in question did not satisfy the assumptions underlying our classification procedure.
Binaries and variables cannot be estimated using
the synthetic grid of models for ``normal'' single stars.  
Exotic stars (from the point of view of 
chemical abundance) like carbon stars were rejected at the very first step
of the method by the method itself, which failed to find satisfying synthetic 
grid points. The same
happens if we try to use photometric observations of a galaxy as an input 
(we used the catalog of Gavazzi \& Boselli,~\cite{gavazzi}). 
In the case of
binaries and variable stars we may get a wrong estimation of the parameters, if we have no idea 
about the true nature of the star and if we have no information about the distance of the star
and the extinction.

\section{Results and conclusions}

\begin{table}[b]
\caption{Reason for the failure of the parameter estimation}
\label{fail_reas}
\begin{tabular}{|c|c|c|c|}
\hline
\multicolumn{4}{|c|}{$\log T_{eff}$} \\
\hline
 name  &  $\log T_{eff}^{sp}$ & $\log T_{eff}^{ph}$ &  \\
\hline
 HD 18078 & 4.003 & 4.654 & A0p \\
 HD 96446 & 4.401 & 4.635 & B2IIIp \\
\hline  
\multicolumn{4}{|c|}{$\log g$} \\
\hline
 name  &  $\log g^{sp}$ & $\log g^{ph}$ &  \\
\hline
  BD +30 2611 & 1.13 & 3.20 & \\
  HD 101602   & 1.40 & 4.88 & UZ Cen, \\
              &      &      & Cep-type variable\\
\hline  
\multicolumn{4}{|c|}{[M/H]} \\
\hline
 name  &  $[M/H]^{sp}$ & $[M/H]^{ph}$ &  \\
\hline
  BD +30 2611 & -1.40 & -0.33 & \\
  HD 101602   &  0.14 & -2.15 & UZ Cen, \\
              &       &       & Cep-type variable \\
  HD 154338   &  0.0  & -1.33 & V991 Sco, \\
              &        &      & RS CVn-type variable\\
  HD 3940     &  -0.39 & -1.85 & V755 Cas, \\
              &        &       & Algol-type \\
  HD 6582     & -0.73  & -1.93 & SB\\ 
\hline
\end{tabular}     
\end{table}

We developed and tested a new method determining the astrophysical parameters from 
broad-band photometry. This method does not rely on a predefined model for the stellar population, 
but requires 
assuming the law for the extinction in the direction of the estimated star. 
In principle, the  extinction law can be determined by the
method itself with the use of stars with a unique solution for their astrophysical parameters, 
which is independent of 
the size of extinction. 

We used an interval-cluster analysis as the mathematical core for the method.  
For a given input photometric system (here Johnson B,V, and 2MASS J, H, and $K_S$), 
this
method allows us to select the regions in the parameter space where a (unique) solution
is possible and to identify regions in the parameter space 
where no solutions are possible due to the complicated topology of the color and parameter spaces.
The astrophysical parameters determined with the use of this method are defined in the system
of the selected models (synthetic or empirical) and suffer from all 
the systematic errors which are inherent in 
these models.
 
We have compiled a catalog of stars with  astrophysical parameters known from spectroscopic observations to
calibrate the Kurucz models and the estimated parameters.

In the Johnson B,V, and 2MASS J, H, and $K_S$ photometric system and with mean errors of 
input colors better than 0.01 mag 
the best results are achieved for main sequence G-K stars (most preferable region, G6 to K9). In this case the parameters are
accurate to $\sigma_{\log T_{eff}} < 0.05$, $\sigma_{\log g} < 0.3$, and $\sigma_{[M/H]} < 0.3$. 
If the accuracy of the
color measurements drops to 0.05 mag, the most preferable region with the accuracy defined above 
($\sigma_{\log T_{eff}} < 0.05$, $\sigma_{\log g} < 0.3$, and  $\sigma_{[M/H]} < 0.3$) 
shrinks to main sequence K6-K9 stars; 
meanwhile, the accuracy for the rest of the most preferable region drops dramatically, especially 
for gravity. Nevertheless, 
even for 0.05 mag the accuracy of the determined parameters for the rest of most preferable regions 
(main sequence G6 to K5) stays 
within $\sigma_{\log T_{eff}} < 0.1$, $\sigma_{\log g} < 2.0$, and $\sigma_{[M/H]} < 0.4$.
The definition of ``the most preferable regions'' depends 
on the combination of photometric bands, desired precision for the result and the precision of the input photometry. 

We find that the biggest problem in determining astrophysical parameters is not to determine a 
solution but to distinguish between possible solutions that correspond to the input colors. 
We propose and tested a simple criterium 
for the selection of a solution based on the estimated extinction. Final tests with 
the compiled catalog proves the reliability and effectiveness 
of the method by comparing spectroscopic (literature) and photometric (our method) determinations of parameters.

The proposed method was  initially designed to search for sparse stellar groups in deep sky
surveys,
but it also suits the study of populations in the Galaxy without a predefined model for the distribution of
stars. It is possible to use the method to determine an extinction law.

\begin{acknowledgements} This research was  
partially performed when ANB was at the Astronomisches Rechen-Institut, Heidelberg, 
where he gratefully acknowledges the DFG grant RO 528/9-1. 
ANB is grateful to Dr. Castelli for help with ATLAS9 program. 
This research  made use of the SIMBAD database, operated at the CDS, Strasbourg, France. 
\end{acknowledgements}

\end{document}